\documentclass{aa}  

\usepackage{graphicx}
\usepackage{txfonts}
\usepackage[normalem]{ulem}
\usepackage{color}

\newcommand{\Fx}{$F_\textrm{x}$} % Flux
\newcommand{\Te}{$T_\textrm{e}$} % electron temperature
\def\ergcms{{\rm erg\,cm^{-2}\,s^{-1}}}

\begin{document} 

   \title{The December 2015 re-brightening of V404 Cyg:}
   \subtitle{Variable absorption from the accretion disc outflow}

   \author{J.~J.~E. Kajava\inst{1}\and
          S.~E. Motta\inst{2}\and
          C. S\'anchez-Fern\'andez\inst{3}\and
          E. Kuulkers\inst{3,4}
          }

   \institute{Finnish Centre for Astronomy with ESO (FINCA), University of Turku, V\"{a}is\"{a}l\"{a}ntie 20, FIN-21500 Piikki\"{o}, Finland\\
              \email{jari.kajava@utu.fi}\and
University of Oxford, Department of Physics, Astrophysics, Denys Wilkinson Building, Keble Road, Oxford OX1 3RH, UK\and
European Space Astronomy Centre (ESA/ESAC), Science Operations Department, E-28691, Villanueva de la Ca\~{n}ada, Madrid, Spain\and
ESA/ESTEC, Keplerlaan 1, 2201 AZ Noordwijk, The Netherlands
}

   \date{Received XXX; accepted YYY}

% \abstract{}{}{}{}{} 
% 5 {} token are mandatory
 
  \abstract
  {In December 2015 the black hole binary V404 Cyg underwent a secondary outburst after the main June 2015 event. 
  We monitored this re-brightening with the \textit{INTEGRAL} and \textit{Swift} satellites, and in this paper we report the results of the time-resolved spectral analysis of these data. 
  The December outburst shared similar characteristics with the June one. 
  The well sampled \textit{INTEGRAL} light curve shows up to 10~Crab flares, which are separated by relatively weak non-flaring emission phases when compared to the June outburst. 
  The spectra are well described by absorbed Comptonization models, with hard photon indices, $\Gamma \lesssim 2$, and significant detections of a high-energy cut-off only during the bright flares. 
This is in contrast to the June outburst, where the Comptonization models gave electron temperatures mostly in the 30--50 keV range, while some spectra were soft ($\Gamma \sim 2.5$) without signs of any spectral cut-off.
Similarly to the June outburst, we see clear sings of a variable local absorber in the soft energy band covered by \textit{Swift}/XRT and \textit{INTEGRAL}/JEM-X, which causes rapid spectral variations observed during the flares. 
During one flare, both \textit{Swift} and \textit{INTEGRAL} captured V404 Cyg in a state where the absorber was nearly Compton thick, $N_\textrm{H} \approx 10^{24}\,\textrm{cm}^{-2}$, and the broad band spectrum was similar to obscured AGN spectra, as seen during the ``plateaus'' in the  June outburst. 
We conclude that the spectral behaviour of V404 Cyg during the December outburst was analogous with the first few days of the June outburst, both having hard X-ray flares that were intermittently influenced by obscuration due to nearly Compton-thick outflows launched from the accretion disc.} 
  

   \maketitle
%
%________________________________________________________________

\section{Introduction}

Most black hole low-mass X-ray binaries show transient behaviour, where they alternate between quiescence and outburst phases (for recent reviews see, \citealt{DGK07,Belloni2016}).
Occasionally weaker secondary outbursts (re-brightenings or re-flares) can follow the initial outbursts by some weeks to a few months (e.g., \citealt{CLG93, BO95, CSL97}).
Analogous secondary outbursts are also seen in systems where the compact object is a white dwarf (e.g., \citealt{KHvP96}) or a neutron star (e.g., \citealt{LRJ10, PMC16}), suggesting that they originate from some type of accretion disc instability, or they are due to an increase in the mass released from the companion star, induced by X-ray heating of its outer layers during the primary outburst (e.g., \citealt{TL95, Lasota2001}).
In 2015 the dynamically confirmed black hole binary V404 Cyg had two bright outbursts, a very bright one in June and a weaker, secondary outburst in December. In this paper we concentrate on the X-ray spectral properties observed during the latter event.

\begin{figure*}
   \centering   
   \includegraphics[width=.56\textwidth, angle=90]{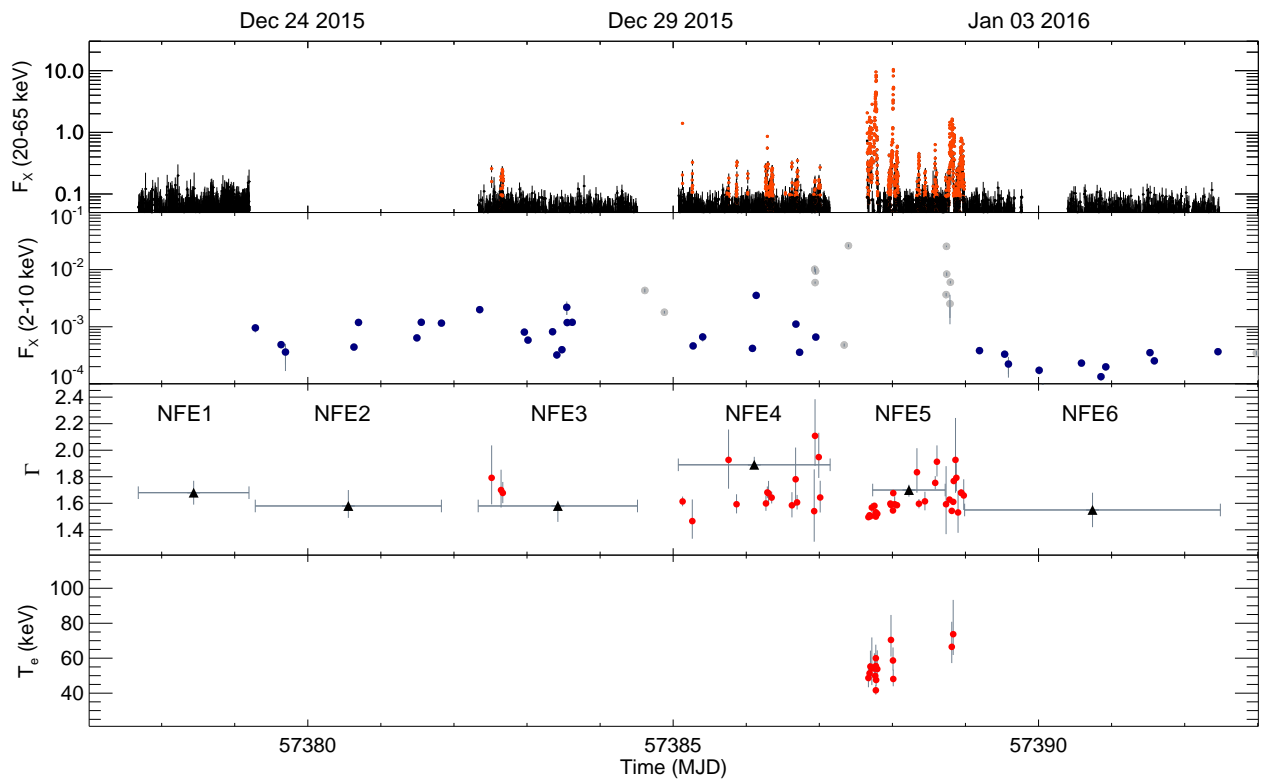}
   \caption{Light curve and spectral parameter evolution of V404 Cyg during the December 2015 outburst. The top panel shows the 20--65~keV \textit{INTEGRAL}/ISGRI light curve, scaled to Crab units. Red symbols are used for the flares. The second panel shows the 2--10~keV fluxes (in Crab units) from the individual \textit{Swift}/XRT snapshots. Blue circles are used for the data that are used to extract the non-flaring spectra. The third panel shows the photon index, with the black symbols marking the spectra accumulated by averaging the \textit{Swift} and \textit{INTEGRAL} non-flaring emission. The bottom panel shows the electron temperatures from the \textsc{nthcomp} model fits.}
    \label{fig:lc}%
\end{figure*}

Similarly to the 1989 X-ray outburst \citep{T89,OvKV96,ZDS99}, during the June 2015 outburst, V404 Cyg showed a spectacular flaring activity, not only in the X-rays (e.g., \citealt{RCBA15,RJB15,JWHH16}), but also in the radio \citep{TSMJ17}, optical \citep{Getal16,KIK16} and in the $\gamma$-rays \citep{LCD16,PMV17}.
X-ray spectral analysis revealed that the flaring was not solely due to mass accretion rate enhancements, but it was strongly influenced by a highly variable local absorber \citep{WMT17, Sanchez-Fernandez2017, MKS17b}.
Occasionally the hydrogen column density exceeded $N_\textrm{H} \approx 1.5 \times 10^{24}\,\textrm{cm}^{-2}$, i.e. the absorber was Compton-thick (the optical depth through the absorber was above unity for Thomson scattering), thus reducing the emission levels throughout the electromagnetic spectrum \citep{MKS17a, Sanchez-Fernandez2017, MKS17b}.
Our interpretation was that the absorption/obscuration was due to a dense and inhomogeneous outflow that was launched from the inner accretion disc, and that the observed flux variability was partly generated when thick clouds (partially) obscured the inner parts of the accretion disc, or when the absorber occasionally became Compton-thin, thus revealing the primary X-ray source.
Further, more direct evidence of winds/outflows was found in high resolution data taken in both the X-ray and optical bands. In \textit{Chandra} grating observations \citet{King2015} found P-Cygni profiles that were indicative of outflow velocities of about a few thousand kilometers per second.
Optical spectroscopy also revealed P-Cygni profiles associated to a few emission lines during the flaring, and the subsequent nebular line emission suggested that significant mass loss had occurred during the outburst \citep{MDCMS16}.

On 23 December, 2015 the \textit{Swift}/BAT telescope detected a re-brightening of V404 Cyg \citep{BPP2015}. Subsequent X-ray, optical and radio observations confirmed the onset of a new outburst \citep{MCM2017} only four months after the end of the primary outburst, which started in June 2015 and reached the X-ray quiescence level early in August 2015 \citep{SBA15,PMJG17}. 
V404 Cyg was already active at least from 21 December, 2015 as \textit{INTEGRAL} serendipitously detected it prior to the BAT trigger (see \citealt{MSF15} and Fig.~\ref{fig:lc}). 
In the December outburst optical observations showed similar signs of winds as during the June outburst, with outflow velocities of about $\sim2500$~km~s$^{-1}$ \citep{MCM2017}.
Both outbursts also showed up to two magnitude optical variability on one hour time scales \citep{KIK16, KKI17}, which were correlated with the X-ray flares.
Furthermore, the radio activity was almost as intense as in June \citep{MCM2017}, suggesting that mass outflows through jets also occurred in both outbursts.  
In this paper, we study the \textit{Swift} and \textit{INTEGRAL} X-ray spectra of V404 Cyg in order to establish whether the Compton thick absorbers that were detected in June play also a role in the December outburst.

\section{INTEGRAL and Swift observations of V404 Cyg during the December 2015 outburst}

The first \textit{INTEGRAL} detection of V404 Cyg during the December re-brightening occurred in the spacecraft revolution 1624, between 21--23 December, 2015 (MJD 57377--57379). After the BAT detection on 23 December, we triggered target of opportunity \textit{INTEGRAL} observations of V404 Cyg that lasted four consecutive revolutions (1626--1629), performed between 25 December, 2015 and 5 January, 2016.
Several \textit{Swift} observations were also triggered, and we analyse here the data that overlap with the \textit{INTEGRAL} monitoring campaign. We indicate the observations and time intervals in Table~\ref{tab:obslog}.
During the last \textit{INTEGRAL} orbit the outburst was seen to fade by \textit{Swift}, and soon after even the \textit{Swift} monitoring had to be stopped due to Sun constraints on 20 January, 2016.
The outburst lasted some time beyond the \textit{Swift} monitoring, as the AMI-LA radio light curve showed significant flaring activity up to 22 January, 2016 \citep{MCM2017}.

\begin{table*}
\centering
\caption{\label{tab:obslog}Log of observations used in this work.} 
\tiny
\renewcommand{\arraystretch}{1.16}
\begin{tabular}{@{}lcccl}
\hline\hline\noalign{\smallskip}
ID                      & \textit{INTEGRAL}                 & \textit{Swift} OBSID                  & Exposures (ksec)                    & Notes                  \\
                         &    revolution              & 000314031XX                 & XRT/JEM-X/ISGRI                    &                  \\

\hline\noalign{\smallskip}
\multicolumn{5}{c}{Non-flare emission (NFE)} \\
\hline\noalign{\smallskip}
NFE1            & 1624       	& ...     & .../49/79          &            \\
NFE2            & ...       	& 21, 22, 25, 26        & 6.5/.../...        &          \\
NFE3            & 1626       	& 27, 28, 23, 24 (snap. 1, 2)        & 7.5/135/106                 &           \\
NFE4            & 1627       	& 29, 31, 32, 00668879000        & 7.9/110/91         & See Fig.~\ref{fig:nonflarespec}          \\
NFE5   		 & 1628       	& ...     & .../47/37                 & \textit{INTEGRAL} data until the last flare \\
NFE6            & 1628, 1629   & 38, 35, 39, 40, 36, 37, 42, 43, 44        & 9.3/.../...        & No \textit{INTEGRAL} detection after the last flare        \\
\hline\noalign{\smallskip}
\multicolumn{5}{c}{Joint \textit{INTEGRAL} and \textit{Swift} flare 1 (JF1)} \\
\hline\noalign{\smallskip}
JF1            & 1627       & 00668877000        & 0.9/0.8/0.6                    & Snapshot 1, first PC mode          \\
\noalign{\smallskip}\hline\noalign{\smallskip}
\multicolumn{5}{c}{Joint \textit{INTEGRAL} and \textit{Swift} flare 2 (JF2); See Figs.~\ref{fig:flarelc}--\ref{fig:broadspectrum}} \\
\hline\noalign{\smallskip}
JF2a            & 1628     &  34       & 0.6/0.4/0.4               & Snapshot 1, first PC mode exposure         \\
JF2b            &         &  34       & 0.1/.../...                & Snapshot 1, WT mode exposure         \\
JF2c            &         &   34      & 0.5/.../...                & Snapshot 1, second PC mode exposure       \\
JF2d            & 1628    &  34       & 0.9/0.6/0.5                & Snapshot 2         \\
\hline 
\end{tabular}
\tablefoot{In the first flare seen jointly by \textit{INTEGRAL} and \textit{Swift} the exposure times in the second snapshot and the WT mode were too short to perform spectral analysis.}
\end{table*}

\subsection{INTEGRAL}

We make use of the public \textit{INTEGRAL} IBIS/ISGRI light curves of V404 Cyg provided by the ISDC \citep{KF16}.
These light curves have a 64~s time resolution, which is appropriate to separate the X-ray flares from the non-flaring emission.
The observed count rate histograms were converted to Crab units with the observed Crab ISGRI rate of 147 cps in the 25--60 keV band.
Based on a close inspection of our dataset, we identified the flares in the ISGRI light curves as follows.
We first estimated the non-flaring emission level as the mean ISGRI count rate during revolution 1624. 
To identify a flare we then \textit{(i)} required that at least 4  time  bins (i.e. 256~s) have a count rate 4.5 $\sigma$ above the non-flaring emission level, but \textit{(ii)}  allowed that during a flare two consecutive points can have a count rate value below the count rate threshold crossed at the onset of each flare, before defining that the flare has ended. 
We then used the obtained time intervals to extract JEM-X and ISGRI spectra.

Since the non-flaring emission is rather faint we integrated over the non-flaring intervals to extract a spectrum in between flares.  We refer to these data as \textit{non-flare emission} (NFE). For the first three INTEGRAL orbits (NFEs 1, 3 and 4) we used the whole \textit{INTEGRAL} orbits. However, for the NFE5 we used only the first half of orbit 1628 where flares were seen, while for NFE6 we used the latter half of orbit 1628 and the whole orbit of 1629, where no flares were seen (see Fig. \ref{fig:lc}). 
During the flares we performed time-resolved spectroscopy any time the number of counts in the flare exceeded a limit of 2$\times$10$^5$ in order to allow a good signal to noise ratio.
If during a flare the total number of counts was lower than the above threshold, we integrated a single averaged spectrum of the flare. 
Conversely, whenever the \textit{INTEGRAL} light curve had gaps (due to space craft slews or perigee passages) we defined that a seemingly continuous flare had ended and started integrating the next flare (see Fig. \ref{fig:flarelc}).

In this manner we extracted 52 ISGRI and JEM-X flare spectra, and 5 non-flaring spectra.
The IBIS/ISGRI and JEMX data reduction was performed using the Off-line Scientific Analysis software (OSA; \citealt{CWB03}) v10.2, using the latest calibration files. 
The data were processed following standard reduction procedures. 
We used good time interval files (GTIs) generated from the flare start/end times calculated as described above to extract the flare spectra. 
We also used GTI files to avoid the flaring intervals in the extraction of the non-flaring spectra, for which we also ignored 300 seconds before and after each flare, to avoid potential flare contamination of the non-flaring intervals.
For the flare spectra we binned the IBIS/ISGRI response matrix in the energy range 20--500\,keV using 28 channels of variable logarithmic widths as in \cite{Sanchez-Fernandez2017}, while for the NFE spectra we used the ISGRI standard 13 channel binning. 
We restricted the spectral fits to the 20--200\,keV energy range, because above 200~keV there is an additional spectral component \citep{JRR17} and the ISGRI data are background dominated.
In our flare fits we have ignored the energy bin around 50\,keV since this data point was systematically below the rest of the spectral bins, and as in \citet{Sanchez-Fernandez2017} we added 3 per cent systematic errors to the spectral bins. 
The JEM-X response matrix in the energy range 3--35\,keV was binned using the standard 8-channel energy bins defined in the OSA software for both flaring and non-flaring spectra. 
In our fits we ignored the first JEM-X energy bin and the noisy channels above 20~keV.

\subsection{Swift/XRT}

We extracted the \textit{Swift}/XRT spectral data and light curves with the on-line Leicester XRT product generator \citep{EBP09}.
To generate the light curve in Crab units, we first converted the \textit{Swift}/XRT count rates to fluxes using WebPIMMS (assuming $\Gamma = 1.7$ and $N_\textrm{H}=0.83\times10^{22}\,\textrm{cm}^{-2}$). This gave a conversion factor of $1\,\textrm{cps} = 5.5\times10^{-11}$ ${\rm erg\,cm^{-2}\,s^{-1}}$ (here the input was 0.3--10 keV band and output was in the 2--10 keV band).
The conversion to Crab units was then done assuming a Crab flux of $2.4\times10^{-8}\,\textrm{erg}\,\textrm{cm}^{-2}\,\textrm{s}^{-1}$.

In most of the observations the count rate was below 1~cps, and thus the data were taken in the photon counting (PC) mode.
In general, observations showing non-flaring emission did not yield many detected photons in \textit{Swift}, and we therefore had to average the non-flare XRT spectra to increase the signal to noise ratio, as indicated in Fig. \ref{fig:lc} and Table \ref{tab:obslog}.
Note that we did not use the XRT data that were taken in between the \textit{INTEGRAL} orbits.
At the times when XRT data were taken during a major flare, some snapshots of a few hundred seconds were sufficient to determine the source spectral shape.
In two cases, also intra-snapshot variability was detected (as highlighted later on). 
In these cases we split the data further either based on the XRT count rate or on the observing mode, as in the flare peaks XRT automatically switched to windowed timing (WT) mode.
All the extracted spectra were binned to have a minimum of 20 counts per channel using the \textsc{grppha} tool.

\subsection{Spectral fits}

We used \textsc{xspec} v12.8.2 to fit the extracted X-ray spectra. 
The best fitting parameters were obtained by minimizing the $\chi^2$ statistics, and the errors provided below are quoted at the 1-$\sigma$ confidence level ($\Delta\chi^2=1$ for one parameter of interest).
We initially fitted the data from the three instruments using freely varying normalization constants, but later found that they were consistent with each other. Therefore, in the spectral fits discussed below these instrument cross-calibration constants are assumed to be unity.

Following our previous analysis of the June 2015 outburst \citep{Sanchez-Fernandez2017}, we used the Comptonization model \textsc{nthcomp} \citep{ZJM96, ZDS99} to fit the spectral data. 
The spectral parameters are the power-law index $\Gamma$ and the electron temperature \Te, with the seed photon temperature fixed to 0.1~keV given that there were no indications of a low-energy cut-off in the XRT data.
To account for the inhomogeneous local absorption that was present in the June outburst \citep{MKS17b}, we multiply \textsc{nthcomp} with a partial covering model \textsc{tbnew\_pcf} and with a second, fully-covering column fixed to the interstellar value of $N_\textrm{H,ISM}=0.83\times10^{22}\,\textrm{cm}^{-2}$ or higher if required by the data using \textsc{tbnew} (obtained from optical reddening of $A_V = 4.0$ \citep{HBR09} and assuming $A_V/N_H (10^{21} \textrm{cm}^{-2}) = 0.48$ \citep{VS15}).
We tested partial covering models on the \textit{INTEGRAL} data, but found that the model parameters could not be constrained without simultaneous \textit{Swift}/XRT data covering the lower energies.
For this reason in the \textit{INTEGRAL}-only fits we fixed the absorption column to the interstellar value.
Similarly, when modelling only the \textit{Swift}/XRT data we used the simpler \textsc{powerlaw} model, since the high-energy cut-off was not visible in the XRT energy band (i.e., below 10~keV). 

After fitting the data with these models we used the method described in \citet{Sanchez-Fernandez2017} to determine if a cut-off, an iron line or if an additional fully or partially covered absorption component were statistically required by the data.
We first approximated the Bayesian information criterion (BIC) as BIC\,=\,$\chi^2 + k \ln(n)$,  
where $k$ is the number of parameters in the model, and $n$ is the number of channels in the spectral fits.
Here a lower BIC implies either fewer explanatory variables, a better fit, or both.
A model with more free parameters was selected if $\Delta$BIC$>$6, i.e. if the strength of the evidence against the model with the higher BIC was strong \citep{KR95}.

In addition to the partially covered direct Comptonization model described above, we also fitted the \textsc{mytorus} model \citep{Murphy2009, Yaqoob2012} to one joint \textit{Swift}/XRT and \textit{INTEGRAL}/JEM-X and IBIS-ISGRI spectrum where large column densities were detected (JF2d). 
This spectral model has been developed to describe obscured AGN where a toroidal absorber reprocesses the central emission in the Compton-thick regime. The model can be adjusted to emulate a toroidal structure, either uniform or patchy, with an half-opening angle of 60$^{\circ}$ and, consequently, a covering fraction of 0.5. We successfully employed this model to describe a \textit{plateau} spectrum during the June 2015 outburst of V404 Cyg \citep{MKS17a}. 
The model we used in the present work consists of a thermal Comptonization component (the \textsc{comptt} \textsc{xspec} model, describing the intrinsic source spectrum), plus two additional components: the \textsc{mytorus} zeroth-order continuum component 
and the \textsc{mytorus} scattered component.
The used model expression in \textsc{xspec} is the following:
\begin{equation*}
\begin{aligned}
\textsc{model} = & \textsc{tbnew}(N_\textrm{H,ISM}) \times(\textsc{constant1} \times \textsc{comptt1}(\tau, T_\textrm{e},N_\textrm{ctt1}) + \\ 
                 & \textsc{comptt2}(\tau, T_\textrm{e},N_\textrm{ctt2}) \times etable\{mytorus\_Ezero\_v00.fits\} + \\ 
                 & \textsc{constant2} \times atable\{mytorus\_scatteredkT050\_v00.fits\})
\end{aligned}
\end{equation*}
Here the optical depths $\tau$ and the model normalizations $N_\textrm{ctt}$ of the two Comptonization components \textsc{comptt1} and \textsc{comptt2} are tied to one another, the electron temperatures are fixed to $T_\textrm{e} = 50$~keV and the seed photon temperature to $T_0 = 0.1$~keV.
The $T_\textrm{e}$ was found by fitting the \textsc{INTEGRAL}/ISGRI spectrum alone above 20 keV using only the zeroth-order continuum component (see \citealt{MKS17a}), under the assumption that such component dominates the high-energies (a reasonable assumption given the high local column density).
The \textsc{mytorus} components both had inclinations fixed to $67\deg$ \citep{KFR10}, the optical depth of the \textsc{mytorus\_scattered} was tied to the \textsc{comptt} component.

The use of first Comptonization component is aimed at describing the intrinsic spectrum of the source, which does not interact at all with the toroidal absorber.
The zeroth-order continuum is formed by the photons that are transmitted through the absorbing torus.
The scattered continuum is, instead, the collection of all escaping photons that have been scattered in the medium at least once. 
This component, in principle, should include both the scattered continuum and the iron line complex. 
However, after a series of tests we found that the data we consider here do not allow to determine whether a patchy reprocessor (requiring two scattered components, with inclination angles fixed at 0$^{\circ}$ and 90$^{\circ}$, respectively) is statistically favoured with respect to a uniform reprocessor.
The fits were neither significantly improved if we included the iron lines to the model.
For these reasons the set-up of our model involves no lines, only one single scattered component and an inclination angle fixed to the source orbital inclination, which implies that the toroidal reprocessor is assumed to be uniform rather than patchy.
In the fitting we changed the abundances to values given by \citet{Anders1989} from the default \citet{Verner1996} ones in order to match the abundances used in the \textsc{mytorus} tables, and we allowed the ISM column density parameter to vary freely.

\section{Results}

The \textit{INTEGRAL}/ISGRI light curve of the December-January outburst is shown in the top panel of Fig.~\ref{fig:lc}.
The X-ray flaring characteristic of V404 Cyg outbursts is clearly visible from MJD 57385 onwards.
The bottom panels of Fig.~\ref{fig:lc} also show the 2--10~keV fluxes of the individual \textit{Swift/XRT} snapshots (in Crab units), as well as the time evolution of the \textsc{nthcomp} model parameters, after separating the flares and the non-flaring emission (see also Tables \ref{tab:nthComp} and \ref{tab:nthCompFlares}). 

\begin{table*}
\centering
\caption{\label{tab:nthComp}Best fitting parameters for fitting the non-flare emission and joint \textit{Swift} and \textit{INTEGRAL} flare data with the following model: {{\sc tbnew} $\times$ {\sc tbnew\_pcf} $\times$ ({\sc nthcomp} $+$ {\sc gauss)}}. The {\sc tbnew\_pcf} and {\sc gauss} models were added or allowed to vary if the improvement to the fit was significant.} 
\tiny
\renewcommand{\arraystretch}{1.16}
\begin{tabular}{@{}lcccccccr}
\hline\hline\noalign{\smallskip}
ID  & $N_{\rm H}$                    & $N_{\rm H,\,pcf}$                 & PCF                   & $\Gamma$                      & $T_{\rm e}$                 & $F_{2-10}$                        & $F_{20-200}$                   & $\chi^{2}/\nu$ \\
  & ($\times 10^{22}$ cm$^{-2}$)                    & ($\times 10^{22}$ cm$^{-2}$)                 &                    &                      & (keV)                 & (cgs)                        & (cgs)                   &  \\
\hline\noalign{\smallskip}
\multicolumn{9}{c}{Non-flare emission (NFE)} \\
\hline\noalign{\smallskip}
NFE1 & [0.83]         & ...       & ...       & $1.68_{-0.09}^{+0.09}$                & $>39$         & $0.054_{-0.032}^{+0.010}$        & $0.188_{-0.036}^{+0.014}$      &  13.3/11 \\
NFE2 & $2.9_{-0.3}^{+0.3}$         & ...       & ...       & $1.58_{-0.12}^{+0.12}$                & ...         & $0.0217_{-0.0012}^{+0.0008}$        & ...      &  63.6/63 \\
NFE3 & $2.27_{-0.13}^{+0.14}$         & ...       & ...       & $1.58_{-0.03}^{+0.04}$                & $>22$         & $0.0266_{-0.0009}^{+0.0009}$        & $0.112_{-0.014}^{+0.015}$      &  95.6/92 \\
NFE4$^{\rm Fe}$ & $1.83_{-0.15}^{+0.2}$         & $26_{-4}^{+4}$       & $0.76_{-0.03}^{+0.03}$       & $1.89_{-0.05}^{+0.06}$                & $>114$         & $0.0574_{-0.002}^{+0.0011}$        & $0.179_{-0.02}^{+0.014}$      &  128.1/107 \\
NFE5 & [0.83]         & ...       & ...       & $1.70_{-0.04}^{+0.05}$                & $34_{-10}^{+30}$         & $0.108_{-0.009}^{+0.008}$        & $0.26_{-0.07}^{+0.02}$      &  8.8/11 \\
NFE6 & $1.27_{-0.2}^{+0.2}$         & ...       & ...       & $1.55_{-0.13}^{+0.13}$                & ...         & $0.0066_{-0.0005}^{+0.0003}$        & ...      &  33.1/41 \\
\hline\noalign{\smallskip}
\multicolumn{9}{c}{Joint \textit{INTEGRAL} and \textit{Swift} flare 1 (JF1)} \\
\hline\noalign{\smallskip}
JF1 & $4.7_{-0.6}^{+0.7}$         & ...       & ...       & $1.38_{-0.07}^{+0.08}$                & $8_{-2}^{+3}$          & $0.21_{-0.12}^{+0.14}$        &  $0.566_{-0.013}^{+0.009}$     &   28.5/34  \\
\noalign{\smallskip}\hline\noalign{\smallskip}
\multicolumn{9}{c}{Joint \textit{INTEGRAL} and \textit{Swift} flare 2 (JF2)} \\
\hline\noalign{\smallskip}
JF2a & [0.83]         & $41_{-7}^{+8}$      & $0.947_{-0.02}^{+0.013}$       & $1.58_{-0.08}^{+0.09}$                & $>30$         & $0.137_{-0.02}^{+0.006}$         & $1.7_{-0.4}^{+0.2}$      &  17.6/21 \\
JF2bm1 & [0.83]         & $4.2_{-0.8}^{+1.0}$       & $0.77_{-0.03}^{+0.03}$       & [1.64]                & ...         & $0.71_{-0.03}^{+0.03}$        & ...      &  64.5/54 \\
JF2bm2 & $1.3_{-0.2}^{+0.2}$          & ...       & ...       & $0.99_{-0.11}^{+0.12}$                & ...         & $0.81_{-0.05}^{+0.04}$        & ...      &  57.8/54 \\
JF2c & [0.83]         & $28_{-7}^{+9}$       & $0.86_{-0.12}^{+0.06}$       & $1.0_{-0.4}^{+0.4}$                & ...        & $0.48_{-0.28}^{+0.04}$        & ...      &  20.6/12 \\
JF2d & [0.83]         & $111_{-8}^{+8}$       & $0.9848_{-0.003}^{+0.003}$       & $1.64_{-0.04}^{+0.04}$                & $>77$         & $0.49_{-0.02}^{+0.02}$        & $12.0_{-0.4}^{+0.3}$      &  23.8/38 \\
\hline 
\end{tabular}
\tablefoot{The Galactic absorption column is fixed to the interstellar value of $0.83\times 10^{22}$~cm$^{-2}$, unless increasing it lead to statistically significant improvement of the fit.
The seed photon temperature was fixed to 0.1 keV.
The fluxes are observed, i.e. not corrected for local nor interstellar absorption and they are given in units of $10^{-9}\,{\rm erg\,cm^{-2}\,s^{-1}}$.
In the ID column $^{\rm Fe}$ notes the detection of an iron line in the NFE4 group: the iron line equivalent width is $EW = 0.31_{-0.09}^{+0.11}$ (line energy and width were fixed to 6.4 and 0.1~keV, respectively). 
}
\end{table*}

\begin{figure}
   \centering
  \includegraphics{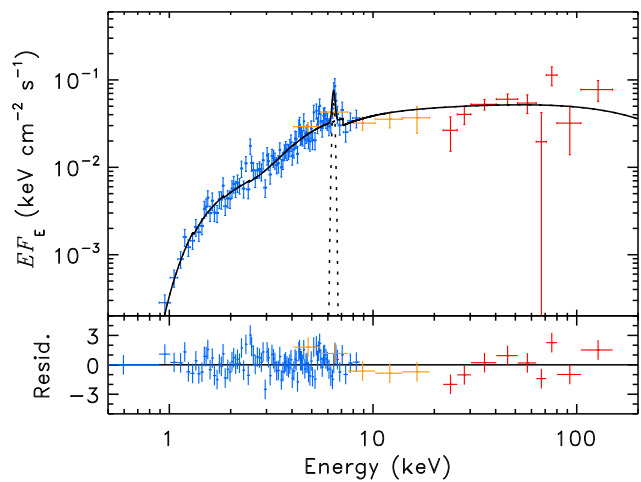}
   \caption{Broad band \textit{Swift}/XRT (blue) and \textit{INTEGRAL} JEM-X (orange) ISGRI (red) spectrum of V404 Cyg of the non-flare emission during MJD 57385--57387 (NFE4). The spectrum is well described by the partially covered Comptonization model, together with a significant, narrow iron line at 6.4~keV. The residuals are in units of sigmas (delchi command in \textsc{xspec}). }
    \label{fig:nonflarespec}%
\end{figure}

An example of a non-flare emission (NFE4) is shown in Fig.~\ref{fig:nonflarespec}.
This non-flare spectrum is the only one where a weak and narrow iron K-$\alpha$ line is seen, with an equivalent width of $EW = 0.31_{-0.09}^{+0.11}$~keV. 
It is also the only spectrum where a good fit requires both a partially covering column and a fully covering column significantly above the interstellar value. 
This spectrum is also the softest of all the non-flare groups, with $\Gamma\approx1.9$ vs. $\Gamma\approx1.6$, despite the fact that  all the NFE spectra have approximately the same emitted flux until the flaring ceased and V404 Cyg became undetectable by \textit{INTEGRAL} in the NFE6 group. 

\begin{figure}
   \centering
   \includegraphics{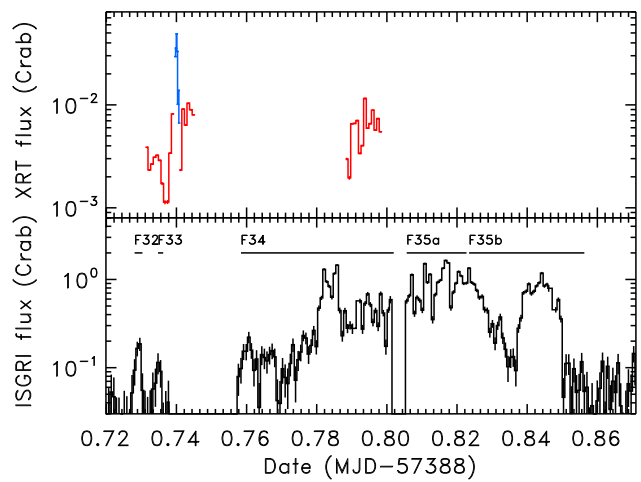}    \caption{Swift/XRT (top) and \textit{INTEGRAL}/ISGRI (bottom) light curves of a flare during MJD 57388.72--57388.87 (Obs. ID 00031403134 from 1 January, 2016). The XRT PC mode data is shown with red and the WT data with blue. Spectral changes during the first XRT snapshot are shown in Fig.~\ref{fig:swiftflare} and the broad band spectrum with \textit{INTEGRAL} and the second \textit{Swift} snapshot is shown in Fig.~\ref{fig:broadspectrum}. The ISGRI-only flare extraction times (FX) tabulated in Table \ref{tab:nthCompFlares} are denoted with horizontal lines. }
    \label{fig:flarelc}%
\end{figure}

During the observing campaign two flares were captured jointly with \textit{Swift} and \textit{INTEGRAL} (denoted as JF1 and JF2 in Tables \ref{tab:obslog} and \ref{tab:nthComp}). 
A zoom-in of the second joint flare is shown in Fig.~\ref{fig:flarelc}.
The top panel shows the \textit{Swift}/XRT light curve (Obs. ID 00031403134), with the PC mode data in red and the WT mode data of a flare peak in blue. 
In the bottom panel a simultaneous \textit{INTEGRAL}/ISGRI light curve is also shown.
Unfortunately, during the strongest XRT flare \textit{INTEGRAL} slewed to the next pointing of its dither pattern, thus missing the action. 

\begin{figure}
   \centering
  \includegraphics{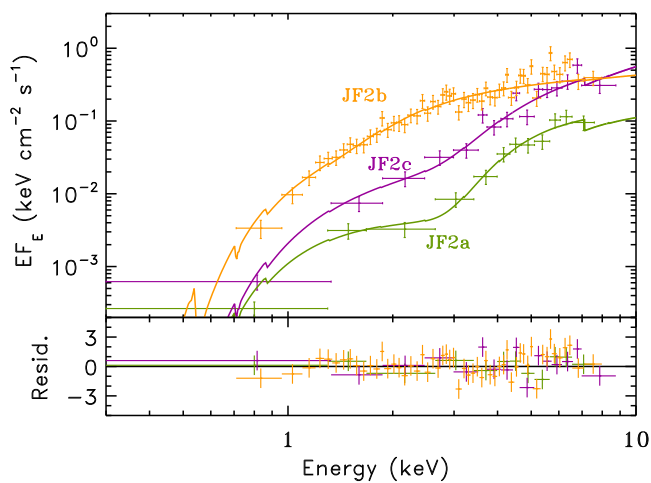}
   \caption{Rapid spectral changes seen during a \textit{Swift} snapshot taken on 1 January 2016 (OBS ID: 00031403134).
   The green spectrum was extracted during the first 590 seconds of the observation (JF2a), the orange spectrum is the WT mode segment covering the 120~s duration flare (JF2b), and the purple spectrum is taken during a 480~s segment posterior to the flare (JF2c). The variability can be well described by a rapid uncovering of the intrinsic emission.
   Similar variability was also seen in the June outburst (see \citealt{MKS17b}, fig. 3).}
    \label{fig:swiftflare}%
\end{figure}

The spectral variations during the first \textit{Swift} snapshot of this flare are shown in Fig.~\ref{fig:swiftflare}.
The first PC mode time interval is shown in green, the WT mode time interval of the flare in orange, and the post flare PC mode interval in purple.
Significant spectral changes are observed in this flare (see also Table \ref{tab:nthComp}).
The best fitting spectral parameters show that in the beginning of the observation (JF2a) the 2--10~keV flux is already enhanced by a factor of two with respect to the non-flaring emission, and that a thick column of $N_{\rm H,\,pcf} = [40\pm7] \times 10^{22}$~cm$^{-2}$ that almost completely covers the source is required.
The subsequent flare -- where the 6-fold increase of flux caused XRT to automatically switch from PC to WT mode -- was instead best fit with only a modest, less covered absorber (JF2bm1), or -- alternatively -- without any partially covering column, but with a significantly harder intrinsic spectrum (JF2bm2).
Neither of these models can be statistically rejected given the relatively noisy XRT data, and without complementary \textit{INTEGRAL} data to constrain the spectral slope.
However, we note that no ISGRI spectra have photon indices below $\Gamma \lesssim 1.4$ (see Tables \ref{tab:nthComp} and \ref{tab:nthCompFlares}), therefore we favour the former best fit.
After the flare the flux drops to a level higher than that seen prior to the flare, and is accompanied by an increase of the partially covering column (JF2c).

\begin{figure*}
   \centering
   \includegraphics[width=0.93\textwidth,angle=0]{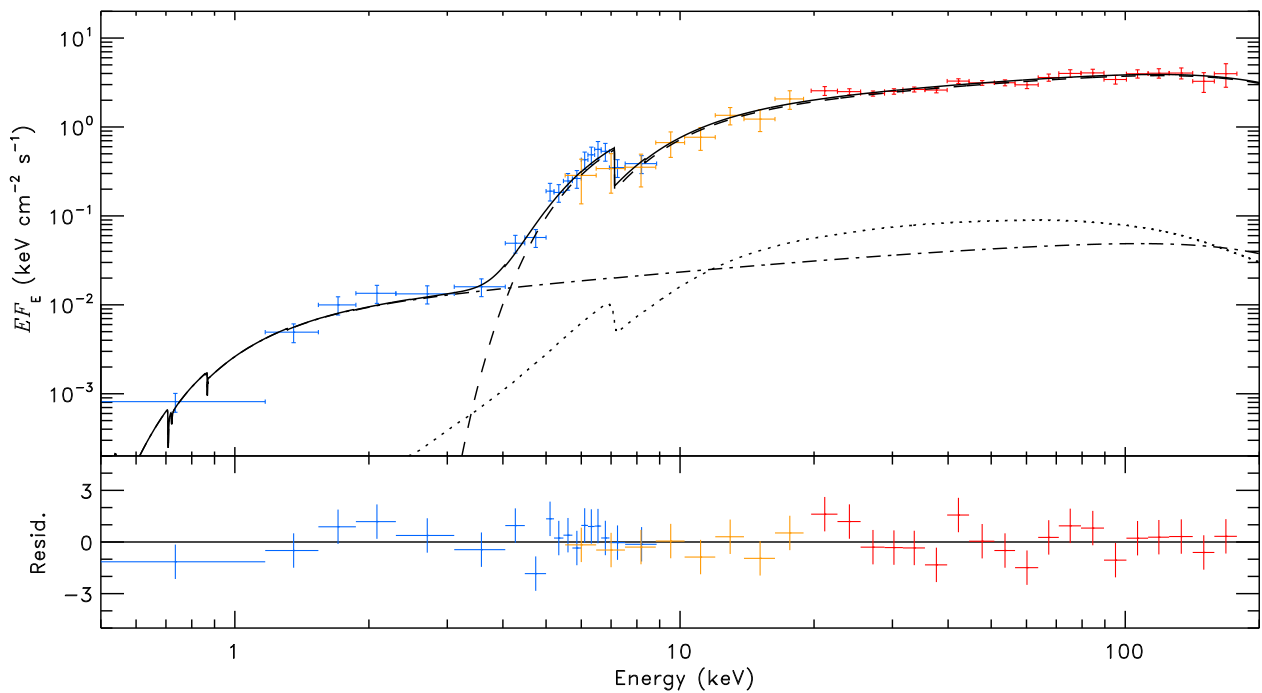}
   \caption{Broad band spectrum of V404 Cyg of a simultaneous \textit{Swift}/XRT (blue) and \textit{INTEGRAL} JEM-X (orange) and ISGRI (red) observation during an X-ray flare JF2d (see Fig.~\ref{fig:flarelc}). The spectrum can be described by a partially covered Comptonized continuum or, as shown here, by the \textsc{mytorus} model \citep{Yaqoob2012}, which describes a thick toroidal reprocessor that scatters the emission of a point-like X-ray source located at the geometrical centre of the torus. The spectral component dominating at high energies is produced by the emission transmitted through a column of $N_\textrm{H} \approx 1 \times 10^{24}\,\textrm{cm}^{-2}$ (dashed line). The dot-dashed line marks the direct Comptonization emission from the source, while the dotted line represents the emission scattered within the toroidal reprocessor before reaching the observer. }
   \label{fig:broadspectrum}%
\end{figure*}

The second \textit{Swift}/XRT snapshot (JF2d) was taken in between two strong, $\sim2$ Crab flares seen the ISGRI data.
We extracted a simultaneous INTEGRAL JEM-X and ISGRI spectra with this XRT snapshot and the resulting broad band spectrum is shown in Fig.~\ref{fig:broadspectrum}.
It can be described well with both the partially covered \textsc{nthcomp} model (see Table \ref{tab:nthComp}), as well as with the \textsc{mytorus} model, which is shown in the figure.
Both models require the presence of a nearly Compton-thick absorber, with $N_{\rm H,\,pcf} \sim 10^{24}$~cm$^{-2}$.
In the \textsc{mytorus} fit the high-energy emission is completely dominated by the so-called transmitted component (or zeroth-order component, see \citealt{Murphy2009, Yaqoob2012, MKS17a}, for details), which accounts for the emission passing through a thick toroidal reprocessor surrounding the source, with a column density of $N_{\rm H,\,pcf} = [1.08 \pm 0.06] \times 10^{24}$~cm$^{-2}$ (dashed line in Fig.~\ref{fig:broadspectrum}). 
The direct Comptonization emission from the central X-ray source, corresponding to the \textsc{comptt1} component in the model description in Section 2.3, that is shown with the dot-dashed line is extremely faint overall, but it dominates the emission at lower energies.
The scattered component (dotted line) is only marginally required to describe the data. 
The \textsc{mytorus} model allows to fit the iron line complex (which constitutes part of the scattered emission) self-consistently with the scattered continuum through a dedicated line component. 
The normalizations of these two components can be tied to one another, so that the line flux is always consistent with the flux of the scattered continuum required by the data. 
However, we find a very low flux of the scattered continuum meaning that the line fluxes are also negligible, which together with the noisy XRT data implied that the line component was not statistically required in the fit.
The best fitting optical depth of the Comptonized continuum is $\tau = 1.27_{-0.03}^{+0.07}$ (electron temperature was fixed to 50~keV).
The observed flux in the 2--10 keV band is $F_{2-10} \approx 4.7\times 10^{-10}\,\ergcms$, but the intrinsic flux measured in the same band was higher by almost a factor of ten, $F_{2-10} \approx 4.2 \times 10^{-9}\,\ergcms$. 
However, in the 20-200 keV band the observed flux was $F_{20-200} \approx 1.2\times 10^{-8}\,\ergcms$, which was only reduced by about 30 per cent as the intrinsic flux in the same band is $F_{20-200} \approx 1.7\times 10^{-8}\,\ergcms$, consistent with the fact that the absorber in this flare was not Compton-thick.

\begin{figure*}
   \centering
   \includegraphics[angle=90]{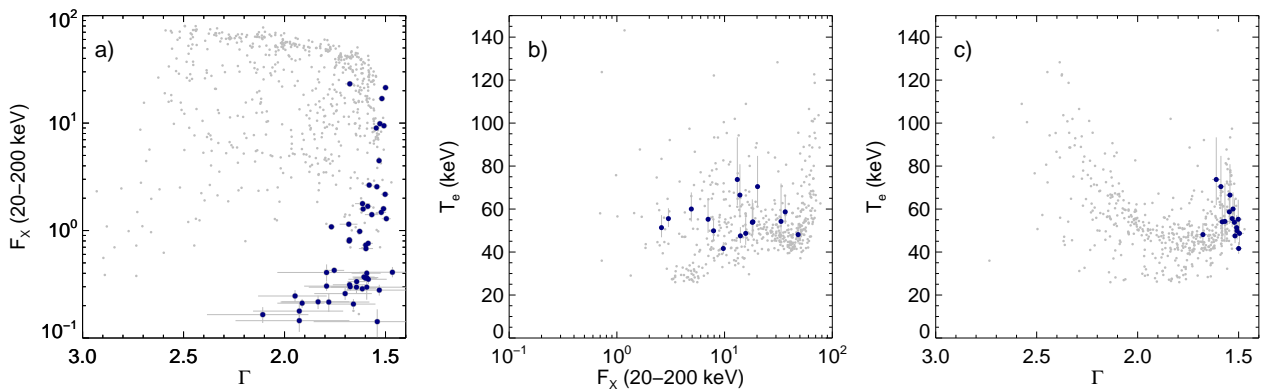}
   \caption{Relations between the spectral parameters from the the {\sc nthcomp} model fits during short and long \textit{INTEGRAL} flares.
   The parameters from the June outburst are also shown using grey symbols.
   In the December outburst, V404 Cyg occupies mostly the ``hard flaring branch'', see panel a). 
   In the \Fx--\Te\ diagram (panel b) we do not see a similar anti-correlation as in the June. 
   Panel c) shows the $\Gamma-$\Te\ diagram. The fluxes are given in units of $10^{-8}\,{\rm erg\,cm^{-2}\,s^{-1}}$.}
              \label{fig:integral}%
\end{figure*}

We identified 39 flares in the ISGRI light curve (see Table \ref{tab:nthCompFlares}).
However, flare 19 (F19) was too faint to perform a meaningful spectral analysis.
Furthermore, in most flares the contemporaneous JEM-X data were very noisy, both due to the high absorption columns and the intrinsic X-ray spectra being so hard.
Additionally, from the \textit{Swift}/XRT data we knew that the absorption model had to be complex, but the JEM-X data alone were not sufficient to constrain the absorption parameters.
For consistency, we therefore decided to consider only the ISGRI spectra for the flares, and used the \textsc{nthcomp} model for these data following \citet{Sanchez-Fernandez2017}.
The relations between the \textsc{nthcomp} model parameters are shown in Fig.~\ref{fig:integral}, together with the results of the June outburst presented in \citet{Sanchez-Fernandez2017} with grey symbols. 
The fits clearly show that during the December outburst V404 Cyg remained in the ``hard flaring branch'' (Fig.~\ref{fig:integral}, panel a), where the spectra are well described by $\Gamma \approx 1.6$ powerlaw.
Only in one bright flare we saw V404 Cyg having a softer $\Gamma \approx 2$ powerlaw spectrum.
We can also see a clear difference between the photon indices of the flare emission below and above $F\approx 5 \times 10^{-9}\,\ergcms$. 
Above this threshold the spectra -- consisting primarily of spectra of long flares -- are clearly harder, with a weighted mean of $\Gamma = 1.619 \pm 0.004$, while below it the spectra of mostly short flares have $\Gamma = 1.75 \pm 0.02$.
In Fig.~\ref{fig:integral}b we do not see a correlation between the electron temperatures with the emitted flux as in the June outburst.
The relations between $T_\textrm{e}$ and $\Gamma$ shown in Fig.~\ref{fig:integral}c cover the same parameter space as during the hard flares seen in beginning of the June outburst, but again, no clear correlation is seen.

\begin{table}
\centering
\caption{\label{tab:nthCompFlares}Best fitting parameters for fitting the ISGRI flare data with the {\sc tbnew} $\times$ {\sc nthcomp} model. Time resolved long ISGRI flares are separated with incrementing alphabets. The flare 19 (F19) was too faint to perform a meaningful spectral analysis.} 
\tiny
\renewcommand{\arraystretch}{1.16}
\begin{tabular}{@{}lcccccr}
\hline\hline\noalign{\smallskip}
ID  & Date           & $\Gamma$           & $T_{\rm e}$          & $F_{20-200}$          & $\chi^{2}/\nu$ \\
    & (MJD)          &                    & (keV)                &     (cgs)             &  \\
\noalign{\smallskip}\hline\noalign{\smallskip}

F1 & 57382.51478            & $1.8_{-0.2}^{+0.2}$   & ...                          & $0.41_{-0.07}^{+0.08}$   & 23.4/17 \\
F2 & 57382.64095            & $1.70_{-0.13}^{+0.15}$   & ...                          & $0.26_{-0.03}^{+0.03}$   & 15.3/17 \\
F3 & 57382.66050            & $1.68_{-0.08}^{+0.08}$   & ...                          & $0.31_{-0.02}^{+0.02}$   & 22.1/17 \\
F4 & 57385.12628            & $1.61_{-0.04}^{+0.04}$   & ...                          & $1.78_{-0.06}^{+0.07}$   & 14.0/17 \\
F5 & 57385.25925            & $1.59_{-0.07}^{+0.08}$   & ...                          & $0.40_{-0.03}^{+0.03}$   & 11.8/17 \\
F6 & 57385.75382            & $1.55_{-0.12}^{+0.14}$   & ...                          & $0.36_{-0.05}^{+0.05}$   & 14.3/17 \\
F7 & 57385.86288            & $1.60_{-0.06}^{+0.06}$   & ...                          & $0.36_{-0.02}^{+0.02}$   & 16.3/17 \\
F8 & 57386.01908            & $1.55_{-0.12}^{+0.14}$   & ...                          & $0.36_{-0.05}^{+0.05}$   & 14.3/17 \\
F9 &  57386.26120           & $1.68_{-0.04}^{+0.05}$   & ...                          & $1.15_{-0.05}^{+0.05}$   & 19.1/17 \\
F10 & 57386.28652           & $1.68_{-0.09}^{+0.09}$   & ...                          & $0.30_{-0.03}^{+0.03}$   & 6.6/17  \\
F11 & 57386.29689           & $1.64_{-0.04}^{+0.05}$   & ...                          & $0.335_{-0.014}^{+0.015}$   & 15.0/17 \\
F12 & 57386.32638           & $1.59_{-0.09}^{+0.10}$   & ...                          & $0.35_{-0.03}^{+0.04}$   & 7.1/17  \\
F13 & 57386.61919           & $1.9_{-0.3}^{+0.3}$   & ...                          & $0.14_{-0.03}^{+0.04}$   & 16.8/17 \\
F14 & 57386.66460           & $1.9_{-0.3}^{+0.3}$   & ...                          & $0.14_{-0.03}^{+0.04}$   & 16.8/17 \\
F15 & 57386.67423           & $1.6_{-0.2}^{+0.2}$   & ...                          & $0.22_{-0.04}^{+0.04}$   & 14.9/17 \\
F16 & 57386.68682           & $1.61_{-0.05}^{+0.06}$   & ...                          & $0.37_{-0.02}^{+0.02}$   & 19.6/17 \\
F17 & 57386.92549           & $1.5_{-0.2}^{+0.3}$   & ...                          & $0.14_{-0.04}^{+0.04}$   & 20.8/17 \\
F18 & 57386.93660           & $2.1_{-0.2}^{+0.3}$   & ...                          & $0.17_{-0.03}^{+0.03}$   & 16.9/17 \\
F20 & 57386.98963           & $1.9_{-0.2}^{+0.2}$   & ...                          & $0.25_{-0.03}^{+0.03}$   & 19.2/17 \\
F21 & 57387.00889           & $1.64_{-0.11}^{+0.13}$   & ...                          & $0.30_{-0.04}^{+0.04}$   & 23.9/17 \\
F22a & 57387.65443          & $1.50_{-0.02}^{+0.02}$   & $49_{-5}^{+7}$    & $1.29_{-0.02}^{+0.02}$  & 8.9/16  \\
F22b & 57387.68924          & $1.50_{-0.02}^{+0.02}$   & $55_{-6}^{+9}$    & $2.17_{-0.03}^{+0.03}$  & 18.6/16 \\
F22c & 57387.71063          & $1.57_{-0.03}^{+0.03}$   & $54_{-10}^{+20}$   & $1.41_{-0.03}^{+0.03}$  & 19.6/16 \\
F23a & 57387.74620          & $1.58_{-0.02}^{+0.02}$   & $54_{-6}^{+9}$    & $2.65_{-0.04}^{+0.04}$  & 8.5/16  \\
F23b & 57387.76239          & $1.51_{-0.02}^{+0.02}$   & $50_{-4}^{+5}$    & $9.44_{-0.10}^{+0.10}$  & 22.0/16 \\
F23c & 57387.76894          & $1.53_{-0.02}^{+0.02}$   & $60_{-6}^{+8}$    & $9.86_{-0.10}^{+0.10 }$  & 17.9/16 \\
F23d & 57387.77413          & $1.50_{-0.02}^{+0.02}$   & $42_{-3}^{+3}$    & $21.4_{-0.2}^{+0.2}$ & 9.0/16  \\
F23e & 57387.77635          & $1.52_{-0.02}^{+0.02}$   & $48_{-3}^{+4}$    & $16.9_{-0.2}^{+0.2}$ & 13.7/16 \\
F23f & 57387.77931          & $1.52_{-0.02}^{+0.02}$   & $54_{-7}^{+11}$   & $1.48_{-0.03}^{+0.03}$  & 19.3/16 \\
F24a & 57387.94133          & $1.60_{-0.02}^{+0.02}$   & ...                          & $0.732_{-0.013}^{+0.013}$   & 19.8/17 \\
F24b & 57388.00629          & $1.55_{-0.02}^{+0.02}$   & $59_{-6}^{+7}$    & $8.98_{-0.09}^{+0.09}$  & 13.4/16 \\
F24c & 57388.01147          & $1.68_{-0.02}^{+0.02}$   & $48_{-4}^{+5}$    & $23.2_{-0.2}^{+0.2}$ & 14.6/16 \\
F24d & 57388.01369          & $1.97_{-0.04}^{+0.04}$   & $39_{-4}^{+6}$    & $10.32_{-0.13}^{+0.13}$  & 12.9/16 \\
F25 & 57388.02459           & $1.59_{-0.06}^{+0.06}$   & ...                          & $0.37_{-0.02}^{+0.02}$   & 9.8/17  \\
F26 & 57388.04829           & $1.59_{-0.02}^{+0.02}$   & ...                          & $0.76_{-0.02}^{+0.02}$   & 12.1/17 \\
F27 & 57388.33224           & $1.83_{-0.15}^{+0.2}$   & ...                          & $0.22_{-0.03}^{+0.03}$   & 10.2/17 \\
F28 & 57388.35505           & $1.60_{-0.04}^{+0.04}$   & ...                          & $0.68_{-0.02}^{+0.02}$   & 18.5/17 \\
F29 & 57388.43725           & $1.62_{-0.07}^{+0.07}$   & ...                          & $0.29_{-0.02}^{+0.02}$   & 12.4/17 \\
F30 & 57388.57621           & $1.75_{-0.05}^{+0.05}$   & ...                          & $0.43_{-0.02}^{+0.02}$   & 22.6/17 \\
F31 & 57388.60018           & $1.91_{-0.11}^{+0.12}$   & ...                          & $0.21_{-0.02}^{+0.02}$   & 15.7/17 \\
F32 & 57387.72808           & $1.81_{-0.14}^{+0.2}$   & ...                          & $0.36_{-0.05}^{+0.05}$   & 13.9/17 \\
F33 & 57388.73475           & $1.6_{-0.2}^{+0.3}$   & ...                          & $0.30_{-0.07}^{+0.08}$   & 21.7/17 \\
F34 & 57388.75840           & $1.63_{-0.02}^{+0.02}$   & ...                          & $0.98_{-0.02}^{+0.02}$   & 11.0/17 \\
F35a & 57388.80567          & $1.54_{-0.02}^{+0.02}$   & $66_{-9}^{+14}$   & $2.56_{-0.03}^{+0.03}$  & 12.7/16 \\
F35b & 57388.82343          & $1.77_{-0.02}^{+0.02}$   & ...                          & $1.08_{-0.02}^{+0.02}$   & 21.0/17 \\
F36 & 57388.86148           & $1.9_{-0.2}^{+0.3}$   & ...                          & $0.15_{-0.03}^{+0.04}$   & 11.8/17 \\
F37 & 57388.87037           & $1.79_{-0.10}^{+0.11}$   & ...                          & $0.30_{-0.03}^{+0.03}$   & 15.6/17 \\
F38 & 57388.88791           & $1.53_{-0.15}^{+0.2}$   & $>23$ & $0.28_{-0.03}^{+0.03}$   & 31.7/16 \\
F39a & 57388.91162          & $1.68_{-0.02}^{+0.02}$   & ...                          & $0.819_{-0.014}^{+0.013}$   & 15.1/17 \\
F39b & 57388.96991          & $1.66_{-0.11}^{+0.12}$   & ...                          & $0.21_{-0.02}^{+0.02}$   & 24.7/17 \\
\hline 
\end{tabular}
\tablefoot{The fluxes are given in units of $10^{-8}\,{\rm erg\,cm^{-2}\,s^{-1}}$.
The seed photon temperature was fixed to 0.1 keV.
}
\end{table}

\section{Discussion}

All but one of the bright flares during the December 2015 outburst of V404 Cyg showed a hard spectrum, with $\Gamma \approx 1.6$, similarly to what is seen in the spectra  obtained during the first few days (i.e. the first \textit{INTEGRAL} orbit) of the June outburst \citep{Sanchez-Fernandez2017}.
Similarly to the June outburst, there is again no need to include thermal disc models in any of the spectra in the soft X-ray band, indicating that the thin disc is likely always truncated and the inner accretion flow consists of some sort of geometrically thick flow.
The rapid flux and spectral variations seen during the flaring shown in Fig.~\ref{fig:flarelc} and Fig.~\ref{fig:swiftflare} were consistent with being due to momentary uncovering of the central source from behind a local, non uniform and nearly Compton-thick neutral absorber.
In addition, the spectrum taken in between two \textit{INTEGRAL}/ISGRI flux peaks shown in Fig.~\ref{fig:broadspectrum} strongly resemble the highly absorbed plateau spectra seen in June \citep{MKS17a,Sanchez-Fernandez2017}.
A narrow emission feature at 6.4~keV consistent with the iron K-$\alpha$ line was detected in one case (NFE4) where thick absorbers were also present, similarly to what was observed in the June outburst.
Hence, we conclude that the flares are again not solely due to accretion events, by their stochastic nature is also affected by local absorption/obscuration of the central source, exactly like in the previous X-ray outbursts from this system \citep{OvKV96,ZDS99,Sanchez-Fernandez2017,MKS17b}.
In a sense these similarities are not surprising, given that the optical spectroscopic data in the December-January \citep{MCM2017} exhibited very similar signatures of outflows as in June-July \citep{MDCMS16}.

One clear difference between the December and June outbursts is seen during the periods in between the flares.
While in June the non-flare emission the 2--10 keV flux was always above $F_{2-10} \gtrsim 10^{-9}\,\ergcms$ \citep{MKS17b}, in the December outburst the non-flare emission was continuously at relatively stable low-flux level at $F_{2-10} \sim [2-10]\times10^{-11}\,\ergcms$. 
These differences are likely not due to stronger absorption/obscuration effects in the December outburst, as the non-flare emission shows only modest local absorption columns. 
Rather, it is more likely that the average accretion rate was  significantly lower in December (which gave a fainter out-of-flare emission) and that the flares resulted from accreting a few dense ``waves'' or clumps, where the accretion rate momentarily increased to near-Eddington levels.\footnote{The 20--200~keV flux during the brightest INTEGRAL flare F24c reached about 20 per cent of the Eddington limit.}

In the June outburst column densities as high as $N_\textrm{H} \sim 3\times10^{24}\,\textrm{cm}^{-2}$ were measured \citep{MKS17b}, being particularly high during the plateau phases in between bright flares \citep{MKS17a,Sanchez-Fernandez2017}.
The analogous event in the December outburst (see Fig.~\ref{fig:broadspectrum}) had roughly 3 times lower column density of $N_\textrm{H} \sim 1\times10^{24}\,\textrm{cm}^{-2}$, which is close to the peak of the high-$N_\textrm{H}$ distribution obtained from the sample of $\sim$1000 \textit{Swift}/XRT spectra taken in June (see fig. 5a in \citealt{MKS17b}).
Consequently, as such absorbers are only nearly Compton-thick, we did not witness the similar factor of ten drops in the total emitted flux during December as we did in June. 
In Fig.~\ref{fig:flarelc} the flux drops in the high-energy ISGRI band were modest, of less than factor of two, consistent with the similar order of magnitude difference between the intrinsic \textsc{mytorus} flux and the observed one.
It is therefore likely that the flux variability was not influenced by the absorption column changes to the same extent as in June.
Nevertheless, the rapid, roughly one minute time scale column variations as seen in the flare shown in Fig. \ref{fig:swiftflare} are very similar to the spectral variations seen in June (see fig. 3 in \citealt{MKS17b}), and thus their origin is likely the same.
As in the June outburst, our interpretation is that during the brightest flares the high accretion rate leads to radiatively driven outflows from the inner thick disc, but that the mass outflow rate was lower given that the December outburst was less intense.

The short flares seen in the December outburst in the flux ranges of $F\sim[1-5]\times10^{-9}\,\ergcms$ were not commonly seen during the June outburst, given that the emission was almost continuously above this range \citep{Sanchez-Fernandez2017}. 
These short flares were significantly softer than the longer, bright flares that had $\Gamma \approx 1.6$.
It is interesting to note that in several other black hole binaries in the hard state -- and also in the AGN population as a whole -- similar spectral softening is observed at comparable luminosity levels \citep{YXY2015}.
These trends are attributed to changes in the amount of external seed photons from the truncated accretion disc that enter the Comptonized region (i.e. the hot flow or corona); 
at luminosities above $L/L_\textrm{Edd} \gtrsim 10^{-3}$ ($F\gtrsim 5\times10^{-9}\,\ergcms$ for V404 Cyg) the disc gradually becomes the dominant source of seed photons for Comptonization as the flux increases, which can explain the observed correlation between $F$ and $\Gamma$ seen in many sources \citep{ZLG03, VVP11, YXY2015, KVT16}.
There are not enough bright flares during the December outburst to see if this trend holds like in the June 2015 outburst, where the brightest flares showed this correlation \citep{Sanchez-Fernandez2017}.

The softening in the $L/L_\textrm{Edd} \lesssim 10^{-3}$ regime ($F\lesssim 5\times10^{-9}\,\ergcms$) instead could be caused by the disc being truncated so far away from the black hole that it plays no role in the Comptonization process, and the observed X-ray spectrum is produced solely via the synchrotron-self-Compton mechanism \citep{YXY2015}.
In this regime the seed photons are generated within the hot-flow, and the $F$--$\Gamma$ anti-correlation could be generated by the accretion flow becoming more and more optically thin as the mass accretion rate decreases (see, e.g., \citealt{VVP11}). Perhaps this type of mechanism was operating in the short flares during the December outburst, giving rise to the softer flares in the $F\sim[1-5]\times10^{-9}\,\ergcms$ flux range.
It has also been speculated that the softening below $L/L_\textrm{Edd} \lesssim 10^{-3}$ is related to the increased contribution of the jet emission to the X-ray band as the flux approaches quiescence (see, e.g., \citealt{SPD11}, and references therein).
However, the analysis of \citet{PMJG17} suggests that the jet does not contribute to the X-ray emission of V404 Cyg even near quiescence levels at $L/L_\textrm{Edd} \sim 10^{-6}$, and it is thus not likely that it plays a role at higher emission levels. 

To conclude, the December 2015 re-brightening can be considered to be a full but ``failed hard state outburst,'' in which the source does not make a state transition to the soft state when approaching the Eddington limit (see, e.g., \citealt{BBF04} and \citealt{KRC16, KVT16, DelSanto2016}, for more recent studies of similar cases).
The spectral variations in V404 Cyg are very similar to other black hole transients, notwithstanding the variable local absorption and the flaring events.
However, during the December outburst V404 Cyg also did not enter the ultraluminous state (i.e. the very high state) despite accreting at near-Eddington levels, as was seen during the major soft flares in the main June 2015 outburst (\citealt{Sanchez-Fernandez2017}) and also for example in the outbursts of GRO J1655--40 \citep{MBvdK98,Motta2012}.
That is, the disc is able to sustain a strong outflow whilst staying in the hard spectral state for at least up to $\sim$20 per cent of the Eddington flux.
In the other flux extreme, we saw in the December outburst that the intrinsic $N_\textrm{H}$ variability occurs at practically all flux levels, also at orders of magnitude below the Eddington limit during the non-flaring emission at $\sim 2 \times 10^{-10}\,\ergcms$. 
This fact is interesting in terms of the launching mechanisms of the outflow. It is hard to imagine that at these lowest fluxes -- more than 1000 times below the brightest flares -- the outflow would only be radiatively driven, as suggested for the brightest flares during the June outburst \citep{MKS17a,Sanchez-Fernandez2017,MKS17b}.
Perhaps some other launching mechanism, such as a mildly relativistic outflowing corona driven by magnetic flares \citep{Beloborodov1999} or a magnetohydrodynamic wind \citep{FKS17}, could be responsible for the $N_\textrm{H}$-variability during these low-flux periods.

\begin{acknowledgements}
We thank the anonymous referee whose comments helped to significantly improve the manuscript. JJEK acknowledges support from the Academy of Finland grant 295114 and the European Space Astronomy Centre (ESAC) for hospitality.
SEM acknowledges support from the Violette and Samuel Glasstone Research Fellowship programme and from the UK Science and Technology Facilities Council (STFC).
SEM and CSF acknowledge support from the ESAC Faculty.
Based on observations with \textit{INTEGRAL}, an ESA project with instruments and science data
centre funded by ESA member states (especially the PI countries: Denmark, France, Germany,
Italy, Switzerland, Spain), and Poland, and with the participation of Russia and the USA.
We acknowledge the use of public data from the \textit{Swift} data archive.

\end{acknowledgements}

\bibliographystyle{aa}

\end{document}